\documentclass[runningheads]{llncs}
\usepackage[T1]{fontenc}
\usepackage{amsmath}
\usepackage{hyperref}
\usepackage{booktabs}
\usepackage{amssymb} 
\usepackage[table,xcdraw]{xcolor} 
\usepackage{multirow}
\usepackage{graphicx,verbatim}
\begin{document}
\title{GNCAF: A GNN-based Neighboring Context Aggregation Framework for Tertiary Lymphoid Structures Semantic Segmentation in WSI}
%

\author{Lei Su\textsuperscript{1} \and Zonghao Liu\textsuperscript{2}\and Lizhi Shao\textsuperscript{3} \and Yang Du\textsuperscript{1}}  
\authorrunning{Lei Su et al.}
\institute{CASMI, Institute of Automation, Chinese Academy of Sciences (CASIA) \and Clinical Oncology School of Fujian Medical University, Fujian Cancer Hospital \and 
School of Internet, Anhui University\\
    \email{sulei2023@ia.ac.cn}, \email{liuzonghao42@gmail.com},\\
    \email{24115@ahu.edu.cn},
    \email{yang.du@ia.ac.cn}
    }

\maketitle              
\begin{abstract}
Tertiary lymphoid structures (TLS) are organized clusters of immune cells, whose maturity and area can be quantified in whole slide image (WSI) for various prognostic tasks.
Existing methods for assessing these characteristics typically rely on cell proxy tasks and require additional post-processing steps.
In this work,
We focus on a novel task-\textbf{T}LS \textbf{S}emantic \textbf{S}egmentation (TLS-SS)-which segments both the regions and maturation stages of TLS in WSI in an end-to-end manner. 
Due to the extensive scale of WSI and patch-based segmentation strategies, TLS-SS necessitates integrating from neighboring patches to guide target patch (\textbf{target}) segmentation. 
Previous techniques often employ on multi-resolution approaches, constraining the capacity to leverage the broader neighboring context while tend to preserve coarse-grained information.
To address this, 
we propose a GNN-based Neighboring
Context Aggregation Framework (GNCAF), which progressively aggregates multi-hop neighboring context from the target and employs a self-attention
mechanism to guide the segmentation of the target.
GNCAF can be integrated with various segmentation models to enhance their ability to perceive contextual information outside of the patch.
We build two TLS-SS datasets, called TCGA-COAD and INHOUSE-PAAD, and make the former (comprising 225 WSIs and 5041
TLSs) publicly available.
Experiments on these datasets demonstrate the superiority of GNCAF, achieving a maximum of 22.08\% and 26.57\% improvement in mF1 and mIoU, respectively.
Additionally, we also validate the task scalability of GNCAF on segmentation of lymph node metastases. 


\end{abstract}
\section{Introduction}

\begin{figure*}[ht]
  \centering
\includegraphics[width=1.0\textwidth]{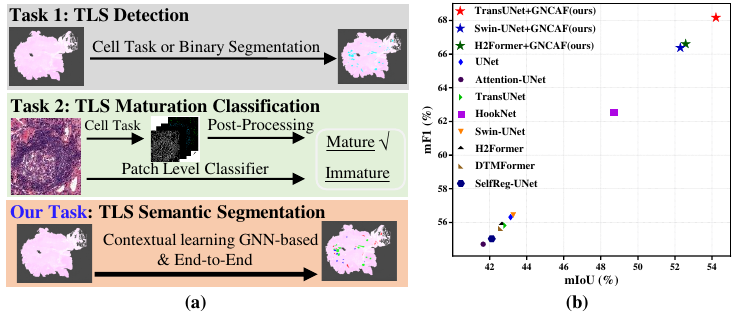} 
  \caption{(a) Comparison of TLS tasks. (b) Comparisons of different method on TCGA-COAD (a TLS-SS dataset we built).}
  \label{fig:intro}
\end{figure*}

Tertiary lymphoid structures (TLS) are organized clusters of immune cells that significantly influence the anti-tumor immune response~\cite{42}.
The maturity and area of TLS are quantified to scores for many tasks such as risk prediction.
To assess these characteristics, pathologists typically mark numerous TLS regions with varying maturity stages in whole slide images (WSI) and sometimes refer to results of multi-immunohistochemistry (m-IHC).
However, the widespread adoption of this approach is limited by time, costs, and available examination techniques. 
In recent years, Computational Pathology (CPath)~\cite{23}
has attracted increasing attention. 
As shown in Fig.~\ref{fig:intro} (a),
the existing tasks of TLS-related on CPath include:
1) \textbf{TLS detection} (TLS-D): Assessing the presence or absence of TLS regions in WSI by markers identification~\cite{li2024machine} or binary segmentation
~\cite{48,49,57,multi_sacle_cnn_04_pixel_level,wang2023weakly,li2024segmentation}, regardless of the maturity of TLS;
2) \textbf{TLS maturation classification} (TLS-MC):
Determining the maturity of TLS through cells proxy task~\cite{46,yang2024cellseg2tls} or classifying the maturity of image patches~\cite{ling2022prognostic}. 
While these tasks hold progress, they are also subject to  several limitations: 1) TLS-D cannot differentiate maturation; 
2) For TLS-MC, cell proxy tasks-based methods require multi-class classification of immune cells and post-processing.
Additionally, methods based on image patch maturation classification only provide patch-level labels, which are inadequate for clinical practice (e.g., the ratio of TLS pixels to the number of tumor pixels).
We focus on a novel task-\textbf{T}LS \textbf{S}emantic \textbf{S}egmentation (TLS-SS) that segments both the regions and maturation stages of TLS in WSI in an end-to-end manner. 
Although this task seems straightforward, many SOTA models show limited performance in TLS-SS, as illustrated in Fig.~\ref{fig:intro} (b).
This arises from the limitations of patch-based segmentation strategies in WSI, particularly the model's inability to 
capture contextual information OUTSIDE the target patch (target) being segmented.
The existing works often rely on multi-resolution methods~\cite{multi_sacle_cnn_01_pixel_level,multi_sacle_cnn_02_pixel_level,multi_sacle_cnn_03_pixel_level,multi_sacle_cnn_04_pixel_level}, 
which limits the capacity and scalability of the model. 
Multi-resolution methods have difficulty capturing the context of a target over a larger field of view and often result in coarse-grained contextual information. 
More effective frameworks for flexibly leveraging context beyond the patch to guide segmentation are still being explored.

\begin{figure*}[ht]
  \centering
\includegraphics[width=0.9\textwidth]{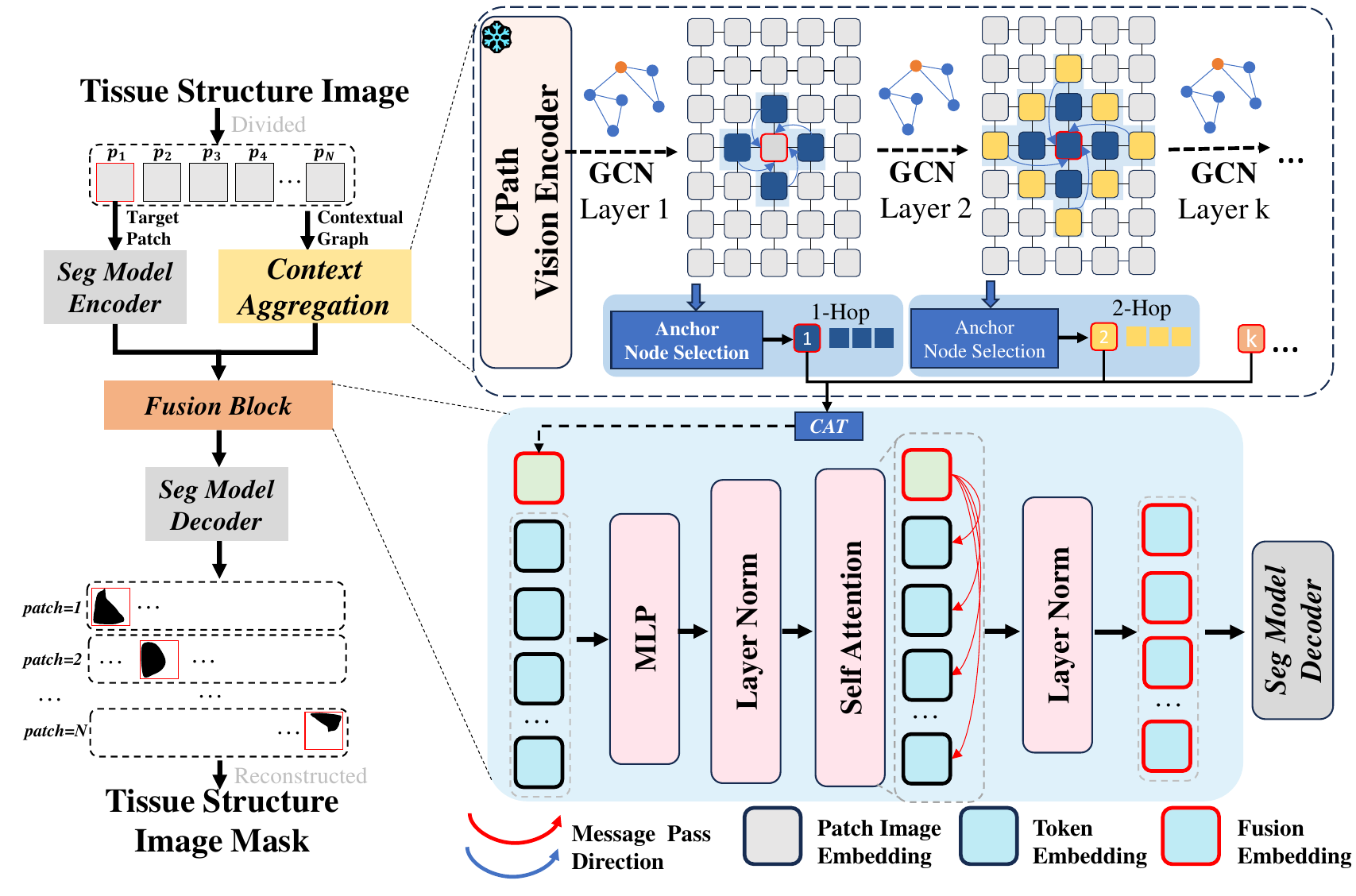} 
  \caption{The architecture of GNCAF.}
  \label{fig:model}
\end{figure*}

Inspired by the ability of Graph Neural Networks (GNN) to flexibly model the context between cells~\cite{contextual_learning_cell_HER2_pre_2022_MIA}, patches~\cite{contextual_gnn_patchgcn_2021_miccai,31}, and other 
interests~\cite{meng2021graph} in medical image, 
a graph can describe the contextual relationships between the target and its neighboring patches in an organized manner (e.g., 4-connectivity and $k$-hops), and GNN can progressively aggregate context from increasingly distant neighbors of the target.
To this end, we propose a \textbf{G}NN-based \textbf{N}eighboring \textbf{C}ontext \textbf{A}ggregation \textbf{F}ramework (GNCAF), which consists of four modules: segment model encoder, segment model decoder, context aggregation module-based GNN, and fusion block.
As illustrated in Fig.~\ref{fig:model},  when given a target to be segmented, the context aggregation module progressively aggregates multi-hop neighboring context of the target, while the fusion block employs a self-attention mechanism to integrate the contextual features with local features for predicting mask. 
GNCAF can be integrated with various segmentation models to enhance the ability of models to perceive contextual information outside of the patch.
To demonstrate the superiority and scalability of GNCAF, we build two TLS-SS datasets (TCGA-COAD and INHOUSE-PAAD) and validate them on an additional task (CAMELYON16\footnote{https://camelyon16.grand-challenge.org/}).
The main contributions of our method include:


1) To the best of our knowledge, we focus on a new task, namely TLS semantic segmentation (TLS-SS) in WSI, which may present new opportunities for WSI segmentation.

2) We present a GNN-based Neighboring Context
Aggregation Framework (GNCAF), which can flexibly aggregate multi-hop neighbor contextual information of the target patch and transmit the context to the target patch features. GNCAF can be easily integrated into various general segmentation models.

3) We build two TLS-SS datasets (TCGA-COAD and INHOUSE-PAAD) from different cancer types for validation. 
Experiments on these datasets demonstrate the superiority and scalability of GNCAF.
Considering the difficulty of acquiring pixel-level annotations in WSI, we release annotated datasets based on TCGA to promote TLS-SS task.

4) We also extend the generalizability of GNCAF in metastases of lymph node metastasis segmentation, demonstrating that it can be applied to other WSI segmentation tasks.

\section{Method}
\subsection{Problem Formulation}
TLS-SS aims to delineate the boundaries of TLS across three maturation stages (e-TLS, pel-TLS, and sel-TLS) in WSI.
As shown in Fig.~\ref{fig:model}, our proposed GNCAF consists of two key blocks: First, the multi-layer GCN iteratively aggregates multi-hop contextual information of target (Section \textbf{Contextual Information Aggregation}). 
Next, the fusion block integrates the context and target features by self-attention for the inferring mask (Section \textbf{Fetures Fusion and Mask Prediction}). The encoder and decoder of the segmentation model can be derived from existing segmentation models.

\subsection{Contextual Information Aggregation}
\textbf{Context Graph Construction}.
To model the neighboring context of the target in WSI, we construct a context graph \( G = (V, E) \)  where \( V \) denotes the set of patch features, and \( E \) represents the set of edges that connect patches.
The foreground of WSI is filtered following \cite{55}, and divided into non-overlapping patches, resulting in a set of \( N \) image patches \( P = \{ p_i \mid i=1 \ldots N \} \).
We use UNI \cite{56} to encode each \( p_i \) into a feature vector \( {v}_i \in \mathbb{R}^{1024} \). 
The undirected edge set \( {E} = \{ {v}_i {v}_j \mid (i,j) \in \mathcal{H} \} \)  for the graph is determined based on the spatial connectivity of patches, where \( \mathcal{H} \) represents the set of naturally connected nodes using 4-connectivity.
The adjacency matrix \( A = [a_{ij}]_{n \times n} \) is derived from the connection relationships between the graph nodes.
Let \( {X}^{(0)} = \{ {x}_1^{(0)}, {x}_2^{(0)}, \dots, {x}_N^{(0)} \} \in \mathbb{R}^{N \times 1024} \) represents the initial feature map for the \( N \) nodes, with each \( {x}_i^{(0)} = {v}_i \).


\textbf{Information Aggregation}.
The aggregation of contextual information outside the patches over $t$ steps can be expressed as:
\[
{X}^{(t)} = \text{F}_{\text{GCN}}({X}^{(t-1)}) = \sigma(\tilde{{A}} {X}^{(t-1)} {W}^{(t-1)}), \tag{1}
\]
where \( \tilde{{A}} = {D}^{-1/2}({A} + {I}){D}^{-1/2} \) represents the normalized adjacency matrix, which is computed to balance the number of neighbors for each node, and \( {D} \) denotes the degree matrix.
From the perspective of a node, the target node \( x_i \) aggregates context from its neighboring nodes, progressively expanding the scope of its contextual awareness.
After \( k \) steps, the representation of node \( i \) is updated from \( {x}_i^{(0)} \) to \( {x}_i^{(k)} \), incorporating contextual information from increasingly distant neighbors.
The set of neighbors within \( t \) steps from $i$-th node
is denoted as \( \text{Nera}_t({x}_i) = \{ {x}_j \mid d(i,j) = t \} \), where \( d(i,j) \) represents the shortest path length between $i$-th node
and $j$-th node.
Therefore, the feature of the $i$-th node  
is updated based on the union of features from all neighbors up to \( K \) steps, which can be expressed as:
\[
{x}_i^{(K)} = \text{F}_{\text{GCN}}( \text{Nera}_t({x}_i^{(K)})),  \tag{2}
\]
This process enables the feature of the node to capture long-range contextual information. 
Therefore, the contextual information of the $k$-hop neighbors can be obtained:
\[
z_i^{context} = \text{MLP}(\text{CAT}([x^{(0)};...;x^{(k)}]) \in \mathbb{R}^{1 \times L} \,  \tag{3}
\]
where \text{CONCAT} represents concatenation in feature dimension and $L$ is the output dimension of the fully connected layer (MLP).
\subsection{Fetures Fusion and Mask Prediction}

We utilize the encoder of the general segmentation model (e.g., TransUNet \cite{50}) to extract the patch features \( \boldsymbol{z}_i^{local} \in \mathbb{R}^{b^2 \times L} \) from target patch \( \boldsymbol{p}_i \in \mathbb{R}^{H \times W \times 3} \), where \( b^2 = \frac{HW}{l^2} \) represents the number of tokens for \( p_i \).
The positional encoding is applied to the patch features \( \boldsymbol{z}_i^{local} \). The two types of features are then concatenated into \( \boldsymbol{z}_i^{(0)} = [\boldsymbol{z}_i^{local} + \boldsymbol{e}_{\text{pos}}; \boldsymbol{z}_i^{context}] \in \mathbb{R}^{(b^2 + 1) \times L} \). 
The fusion block consists of \( \ell \) layers of multi-head attention (MSA). The final fused feature is computed as follows:
\begin{equation}
\boldsymbol{z}_i^{(\ell)} = \text{MSA}^{(\ell)}(\text{LayerNorm}({z}_i^{(0)}))
\tag{4}
\end{equation}
where
\( \boldsymbol{z}_i^{(\ell)} \in \mathbb{R}^{(b^2 + 1) \times L} \) is the output of the feature fusion. 
We select the features corresponding to the token positions, denoted as \( \boldsymbol{z}_i^{\prime (\ell)} \in \mathbb{R}^{b^2 \times L} \). These features are then fed into the segmentation model decoder to predict the mask \( \boldsymbol{y}_m^{\text{pred}} \in \mathbb{R}^{c \times H \times W} \), where $c$ refers to the predefined number of categories for the segmentation targets.
The segmentation loss is computed using cross-entropy and the network is optimized in an end-to-end way.

%
%
%
%
\section{Experiments}
\begin{table}[ht]
\caption{Comparison with SOTA models on two datasets. \textbf{Bold} indicates the best performance, and uparrow indicate the best improvement.}
\label{tab:sota}
\centering
\begin{tabular}{lcccccccc}
\toprule
\multicolumn{1}{c}{} & \multicolumn{4}{c}{TCGA-COAD} & \multicolumn{4}{c}{INHOUSE-PAAD} \\
\cmidrule(r){2-5} \cmidrule(l){6-9}
 Method& mF1 & mIoU & mP & mR & mF1 & mIoU & mP & mR \\
\midrule
UNet\cite{50} & 56.30 & 43.13 & 56.58 & 56.27 & 55.47 & 41.69 & 56.63 & 54.99 \\
Attention-UNet\cite{compared_attention_unet} & 54.71 & 41.65 & 55.23 & 55.63 & 52.71 & 39.82 & 53.70 & 53.20 \\
HookNet\cite{multi_sacle_cnn_04_pixel_level}  & 62.55 & 48.68 & 63.13 & 62.65 & 60.36 & 45.96 & 60.39 & 60.36 \\
DTMFormer\cite{compared_dtmformer} & 55.64 & 42.59 & 57.70 & 54.42 & 53.91 & 40.33 & 54.60 & 54.72 \\
SelfReg-Net\cite{zhu2024selfreg} & 55.03 & 42.11 & 55.22 & 56.08 & 54.45 & 40.81 & 54.52 & 54.89 \\

TransUNet\cite{compared_attention_transunet} & 55.84 & 42.83 & 59.79 & 54.24 & 54.09 & 40.45 & 54.57 & 54.48 \\ 
\rowcolor{gray!15}\phantom{N}\textbf{+GNCAF} & \textbf{68.17\textcolor[rgb]{1.0,0.0,0.0}{$\uparrow$}} & \textbf{54.21\textcolor[rgb]{1.0,0.0,0.0}{$\uparrow$}} & \textbf{68.12} & \textbf{68.56\textcolor[rgb]{1.0,0.0,0.0}{$\uparrow$}} & 62.67\textcolor[rgb]{1.0,0.0,0.0}{$\uparrow$} & 48.06\textcolor[rgb]{1.0,0.0,0.0}{$\uparrow$} & \textbf{64.42}\textcolor[rgb]{1.0,0.0,0.0}{$\uparrow$} & 62.11\textcolor[rgb]{1.0,0.0,0.0}{$\uparrow$} \\
Swin-UNet\cite{compared_attention_swinunet} & 56.44 & 43.27 & 57.75 & 55.38 & 56.31 & 42.38 & 57.04 & 56.61 \\
\rowcolor{gray!15}\phantom{N}\textbf{+GNCAF} & 66.38 & 52.29 & 68.09 & 64.88 & 59.93 & 45.15 & 60.82 & 59.75 \\
H2Former\cite{compared_h2former} & 55.87 & 42.67 & 56.49 & 55.60 & 55.77 & 41.91 & 56.09 & 56.60 \\
\rowcolor{gray!15}\phantom{N}\textbf{+GNCAF} & 66.60 & 52.58 & 67.34\textcolor[rgb]{1.0,0.0,0.0}{$\uparrow$} & 65.91 & \textbf{62.84} & \textbf{48.19}& 64.19 & \textbf{62.56} \\
\midrule
Best Improvement (\%) & 22.08 & 26.57 & 19.20 & 26.40 & 15.86 & 18.81 & 18.05 & 14.00\\
\bottomrule
\end{tabular}
\end{table} 
\subsection{Datasets and Implementation Details}
We build two TLS-SS datasets (TCGA-COAD and INHOUSE-PAAD)
from the colon adenocarcinoma and the pancreatic adenocarcinoma for validation.
TCGA-COAD was downloaded from TCGA\footnote{https://www.cancer.gov/ccg/research/genome-sequencing/tcga} and cleaned by excluding low-quality WSIs. 
There are 225 WSIs annotated by pathologists, without the assistance of m-IHC, including 3496 e-TLS, 1034 pel-TLS, and 511 sel-TLS. 
The INHOUSE-PAAD dataset is from a private hospital. Pathologists annotated TLS in 108 WSIs aided by m-IHC, including 2,339 e-TLS, 1,586 pel-TLS, and 611 sel-TLS.
For each dataset, the training, validation, and testing sets were randomly divided in a 6:2:2 ratio at WSI-level.

The patch size is set to 256$\times$256 with a spatial resolution of 1 $\mu$m/pixel. 
Softmax is employed in the aggregation function of the GCN layer. The number of neighboring hops for aggregation is set to 3.
The number of layers in MSA is 1, with 8 heads. For TransUNet~\cite{compared_attention_transunet}, we utilized the attention layers of model for feature fusion.
During training, the batch size is set to 16, and the learning rate is set to $5\times10^{-5}$. 
We report the mF1, mIoU, mP, and mR to evaluate overall segmentation performance.

\begin{figure*}[ht]
  \centering
\includegraphics[width=0.9\textwidth]{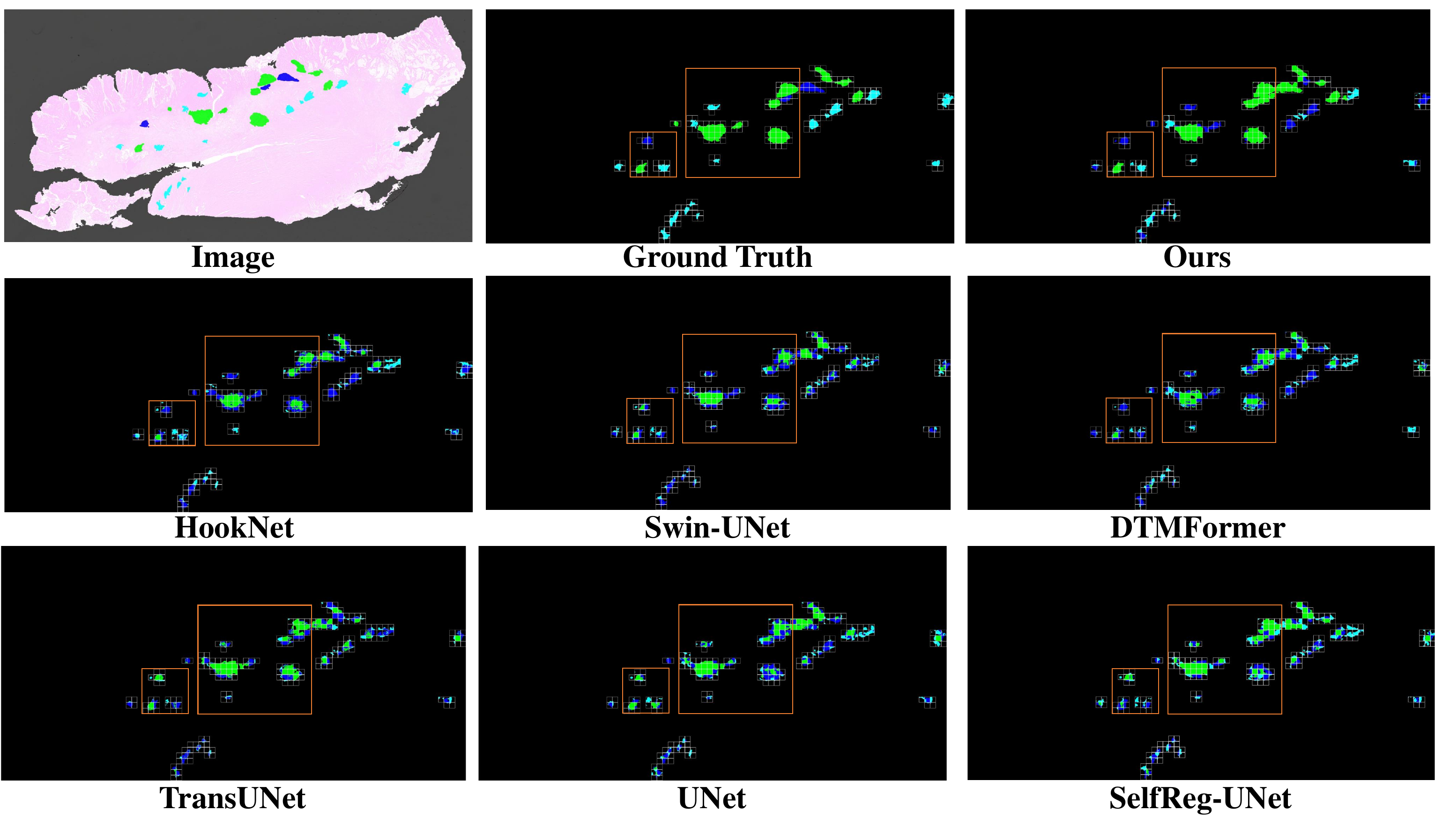} 
  \caption{The visualization of different models on TCGA-COAD. 
  Light blue: e-tls. Bule: pel-TLS. Green: sel-TLS. Ours: TransUNet+GNCAF.}
  \label{fig:vision}
\end{figure*}

\subsection{Comparisons with State-of-the-Art Methods}
As shown in Table~\ref{tab:sota},  we combine three segmentation models with our framework. GNCAF significantly improves the performance of segmentation models.
\textbf{TransUNet~\cite{compared_attention_transunet}$+$GNCAF} achieves the highest performance on TCGA-COAD, with improvements of 22.08\% in mF1 and 26.57\% in mIoU. 
On INHOUSE-PAAD, \textbf{H2Former~\cite{compared_h2former}+GNCAF} achieves the highest mF1 and mIoU, with values of 62.84\% and 48.19\%, respectively.
 This demonstrates that our framework effectively leverages contextual information to guide the target patch toward more accurate segmentation. TransUNet~\cite{compared_attention_transunet} is selected as the segmentation model for subsequent validation. 
 
Visualization presented in Fig.~\ref{fig:vision} demonstrate that our method achieves close alignment with the ground truth, especially in preserving the integrity of TLS regions.
Other models struggle to effectively utilize contextual information due to their inability to fully leverage information beyond the patch, resulting in weaker intra-region consistency.

\begin{table}[ht]
\caption{Ablation studies of three components of GNCAF on two Datasets.}
\label{tab:ablation}
\small
\centering
\begin{tabular}{lcccccccc}
\toprule
\multicolumn{1}{c}{}&\multicolumn{3}{c}
 {Ablation Setup} & \multicolumn{2}{c}{TCGA-COAD} & \multicolumn{2}{c}{INHOUSE-PAAD} \\
\cmidrule(r){2-4} \cmidrule(l){5-6} \cmidrule(l){7-8}
Group & UNI\cite{56} & GCN & MSA & mF1(\%) & mIoU(\%) &mF1(\%) & mIoU(\%) \\
\midrule
$\mathbb{A}$ & ResNet (frozen)~\cite{he2016deep} & $\checkmark$ & $\checkmark$ & 56.40 & 43.12 & 55.99 & 42.07 \\
$\mathbb{B}$ & ResNet (finetune)~\cite{he2016deep} & $\checkmark$ & $\checkmark$ & 63.19 & 49.20 & 59.24 & 44.87 \\
$\mathbb{C}$ & CONCH (frozen)~\cite{lu2024visual} & $\checkmark$ & $\checkmark$ & 63.94 & 49.87 & \textbf{64.38} & \textbf{49.63} \\
$\mathbb{D}$ & $\checkmark$ & ADD & $\checkmark$ & 63.12 & 49.16 & 56.36 & 42.45 \\
$\mathbb{E}$ & $\checkmark$ & MEAN & $\checkmark$ & 65.04 & 50.95 & 55.73 & 41.59 \\
$\mathbb{F}$ & $\checkmark$ & MSA & $\checkmark$ & 67.86 & 53.99 & 61.03 & 46.54 \\
$\mathbb{G}$ & $\checkmark$ & $\checkmark$ & CAT & 65.10 & 50.82 & 58.61 & 44.25 \\
$\mathbb{H}$ & $\checkmark$ & $\checkmark$ & DOT & 65.53 & 51.44 & 61.06 & 46.22 \\
$\mathbb{I}$ & $\checkmark$ & $\checkmark$ & $\checkmark$ & \textbf{68.17} & \textbf{54.21} & 62.67 & 48.06 \\
\bottomrule
\end{tabular}
\end{table}


\begin{table}[ht]
\caption{Model performance (\%) of different hops of neighboring context. Segment model: TransUNet.}
\label{tab:hops}
\centering
\begin{tabular}{lccccccccc}
\toprule
Hops & 0 & 1 & 2 & 3 & 4 & 5 & 6 & 7\\
\midrule
TCGA-COAD   & 55.84          & 66.04 & 67.04         & 68.17               & 68.10         & \textbf{69.90}         & 69.87 & 68.91\\
INHOUSE-PAAD     & 54.09       & 59.92  &59.10    & \textbf{62.67 }              & 61.25         & 61.68         & 62.36 &60.08\\
\bottomrule
\end{tabular}
\end{table}
\subsection{Ablation Study}
To assess the effectiveness of GNCAF, we conduct ablation experiments on Patch Encoding ($\mathbb{ABC}$, default UNI~\cite{56}), Context Aggregation ($\mathbb{DEF}$, default GCN), and Fusion Block($\mathbb{GH}$, default MSA).
\textbf{Patch Encoding}. 
The pre-trained ResNet18~\cite{he2016deep} underperformed, while fine-tuning it in $\mathbb{B}$ significantly improved performance; the vision-language model CONCH~\cite{lu2024visual} achieved the best results. These findings demonstrate that training the backbone or adopting a CPath foundation model is reliable, though directly training the encoding network incurs additional computational overhead.
\textbf{Context Aggregation}.
The utilization of elementary fusion approaches ($\mathbb{D}$ and $\mathbb{F}$) leads to notable performance deterioration, while the self-attention mechanism exhibits inherent stability limitations.
This indicates that GCN is more effective for aggregating context information of the target patch. 
\textbf{Fusion Block}.
We tested two basic fusion methods: concatenation (CAT) and dot product (DOT). Both underperformed, demonstrating the limited efficacy of simplistic fusion in our framework. In contrast, the attention mechanism dynamically injects contextual features into target patches, significantly boosting segmentation accuracy.
\textbf{Hops of Neighboring Context}.
As shown in Table~\ref{tab:hops}, reducing the context aggregation hop distance severely degrades segmentation performance, with the lowest metrics observed in zero-context scenarios. However, when extending the hop distance beyond an optimal range, performance plateaus after a minor dip.
These findings collectively suggest that the hop parameter serves as a critical indicator for context aggregation, and its optimal value may require dataset-specific calibration.
\textbf{Task Generalization}. To validate the extensibility of GNCAF, we adapted our framework to lymph node metastasis segmentation using in Camelyon16. As shown in Table~\ref{tab:cam16}, our framework achieved comparable performance (90.01\% mF1) to many SOTA models. This demonstrates potential of GNCAF for generalizing to other WSI segmentation tasks, where its context-aware architecture can boost performance through neighboring context
aggregation.

\begin{table}[ht]
\caption{Additional analysis on Cameron 16 dataset. Ours:TransUNet+GNCAF.}
\label{tab:cam16}
\centering
\begin{tabular}{lccccccccc}
\toprule
\multirow{2}{*}{\textbf{}} &  &  & Attention- & DTM & H2 & Hook & Swin- & SelfReg- & Trans \\
&  Ours  & U-Net & UNet        & Former & Former & Net & UNet & UNet & UNet \\
\midrule
mF1(\%)             & \textbf{90.01} & 85.23         & 84.03               & 87.17         & 86.32         & 86.09         & 85.08         & 86.13          & 86.72         \\
mIoU(\%)             & \textbf{82.16} & 74.90         & 73.01               & 77.76         & 76.49         & 76.10         & 74.60         & 76.18          & 77.03         \\
\bottomrule
\end{tabular}
\end{table}
\section{Conclusion}
In this paper, we introduce a novel task of TLS Semantic Segmentation (TLS-SS) and propose a GNN-based Neighboring Context Aggregation Framework (GNCAF) to progressively leverage multi-hops neighboring context to guide target patch segmentation. 
The effectiveness and scalability of GNCAF are demostrated on two self-built TLS-SS datasets and an additional lymph node metastasis segmentation task. To promote TLS-SS, We release an annotated datasets based on TCGA-COAD (225 WSIs and 5041 TLSs).

\bibliographystyle{splncs04}
\bibliography{main}
\end{document}